\documentclass{sig-alternate}
\usepackage{}
\usepackage{multirow}
\usepackage{mathrsfs}
\usepackage{graphicx}  
\usepackage{float}
\usepackage{amssymb}
\usepackage{amsmath}
\usepackage{cases}
\usepackage{amsfonts}
\usepackage{mathrsfs}
\usepackage[linesnumbered,ruled]{algorithm2e}

\newtheorem{theorem}{Theorem}[section]

\newtheorem{corollary}[theorem]{Corollary}
\newtheorem{lemma}[theorem]{Lemma}

\newtheorem{define}[theorem]{Definition}

\def\qed{\hfil {\vrule height5pt width2pt depth2pt}}
\def\wrt{w.r.t}

\def\qed{\hfil {\vrule height5pt width2pt depth2pt}}

\renewcommand{\thefigure}{\arabic{figure}}

\def\qed{\hfil {\vrule height5pt width2pt depth2pt}}
\def\S{\mathcal{S}}
\def\C{\mathcal{C}}

\def\wrt{w.r.t.\,}

\def\proof{\textbf{Proof. }}

\def\Res{\hbox{\rm{Res}}}
\def\res{\hbox{\rm{Res}}}

\def\OO{\mathcal {O}}

\def\Q{{\mathbb Q}}
\def\R{{\mathbb R}}

\def\CC{{\mathbb C}}
\def\V{{\mathbb V}}

\def\O{\tilde{\mathcal {O}}}

\def\LL{{\mathcal L}}

\def\Z{{\mathbb{Z}}}

\def\G{{\mathscr{G}}}

\def\P{{\mathscr{P}}}

\begin{document}

\title{On the Complexity of Computing the Topology of Real Algebraic Space Curves}

\numberofauthors{1}

\author{\small Kai Jin$^1$,  \hspace{0.3cm} Jin-San Cheng$^2$\\
\small $^1$School of Mathematics and Statistics $\&$ Hubei Key Laboratory of Mathematical Sciences, \\
\small Central China Normal University, Wuhan, China, 430079.\\
 \small $^2$ KLMM, Academy of Mathematics and Systems Science, Chinese Academy of Sciences,Beijing, China 100190\\
{\small jinkaijl@163.com, jcheng@amss.ac.cn}}

\maketitle
\begin{abstract}
In this paper, we present a deterministic algorithm to find a strong generic position for an algebraic space curve. We modify our existing algorithm for computing the topology of an algebraic space curve and analyze the bit complexity of the algorithm. It is $\tilde{\mathcal {O}} (N^{20})$, where $N=\max\{d,\tau\}$, $d, \tau$ are the degree bound and the bit size bound of the coefficients of the defining polynomials of the algebraic space curve. To our knowledge, this is the best bound among the existing work. It gains the existing results at least $N^2$.

\end{abstract}




\section{Introduction}
Algebraic space curves are used in computer aided (geometric) design, and geometric modeling. Computing the topology of an algebraic curve is also a basic step to compute the topology of algebraic surfaces \cite{chengtop,Diatta}. There have been many papers studied the guaranteed topology and meshing for plane algebraic curves \cite{amj,D-arnon,Berberich11,yap-issac2008,top-cur,eigenwillig08,eigenwillig,hong,vega,Seidel2005}. But there are only a few papers which studied the guaranteed topology of space algebraic curves \cite{JGA05,NSPC,SPC,dia,mourrainsc,ElKahoui08}. The complexity analysis for computing the topology of an algebraic space curve is also not deeply studied. In this paper, we will deal with this problem, which is one contribution of the paper.

Most of the existing work (\cite{JGA05,NSPC,SPC,dia,ElKahoui08}) of computing the topology of algebraic space curves
require the space curve to be in a generic position. But checking whether an algebraic space curve is in a generic position or not is not a trivial task, see \cite{JGA05,SPC,Diatta}. In this paper, we will give a deterministic algorithm to find a generic position for an algebraic space curve, which is another contribution of the paper.

\noindent {\bf Related work}  In \cite{JGA05}, the space curve is assumed to be without singular points and in a generic position related to both the $xy$-plane and $xz$-plane, and any point of the algebraic space curve will not correspond to a singularity of both projection curves. Their generic position checking involves mainly resultant computation. In \cite{ElKahoui08}, ElKahoui considers the algebraic space curve defined by the $n$ tri-variate polynomials. They give a generic position definition (See Definition 4.1 in \cite{ElKahoui08}) which is stronger than Definition \ref{GenericPosition} in this paper and provide a method to check it, which involves computation of Gr\"{o}bner basis. 
In \cite{dia}, the authors require the space curve to be in a generic position and any apparent singularities should be a node. They check the generic position by computing the subresultant sequence related to the defining polynomials of the space curve. The checking of another condition involves testing whether the Hessian matrix of the projection curve of the algebraic space curve at a plane algebraic point is regular or not. In \cite{SPC}, it requires only the algebraic space curve to be in a weak generic position. In \cite{NSPC} the authors provide a deterministic and  easy way to compute a weak generic position for an algebraic space curve.


For the complexity of computing the topology of algebraic space curves, Diatta et.~al
\cite{Diatta} give a descriptive bit complexity of $\tilde{\mathcal
{O}}(d^{21}\tau)$ for the topology computation of an algebraic space curve implied
the assumption of that the input algebraic space curve is in a generic
position. Cheng et.~al \cite{SPC} give a complexity analysis
which is bounded by $\tilde{\mathcal {O}}(d^{37}\tau)$ also implied the assumption that the space curve is in a weak generic position hypothesis. Both of the algorithms imply the assumption of generic position since they provide only the way to check the generic position but not to find a generic position for the algebraic space curve.

\noindent{\bf Our contributions} The main contributions of this paper are as follows. Firstly, we find a method which deterministically puts the original algebraic space curve in a generic position. It is easy to be implemented. As we know, all of the methods mentioned above except \cite{NSPC} provide only a way to check the generality of an algebraic space curve, they do not present a method to find a certified generic position deterministically. Secondly, we present an algorithm to compute the topology of an algebraic space curve, which is modified from the algorithm in \cite{NSPC}. Explicitly, we mainly modified the lifting step, such that the number of space point candidates are bounded by $d^2$ not $d^3$ on each $x$-fiber. Actually, this small change improves the complexity of lifting process by a factor of $d^2$, which leads to a smaller bound for the total complexity of the algorithm. The algorithm presented here is only for bit complexity analysis. There is no implementation. Finally, we analysis the bit complexity $\O (d^{19}(d+\tau))$ of computing the topology of any given algebraic space curve (without any assumptions). 

The paper is organized as follows. In the next section, we give some notations and preliminaries. In the section 3, we present a method to compute a deterministic generic position. The algorithm to compute topology of an algebraic space curve is presented in section 4. We analyze the bit complexity of the algorithm in the last section.

\section{Notations and preliminaries}
\label{Notations}
In this section, notations and known results needed in this paper
are given. Let $\Q,\R,\mathbb{C}$ be the fields of rational numbers,
real numbers and complex numbers respectively. $\Z$ is the set of
integers.

For the simplicity of notations, we use $[q],\,q\in\Z_{+}$ to denote the set of positive integers not greater than $q$.

Let $h(x,y)\in \mathbb{R}[x,y]$. We denote the plane algebraic curve defined by $h=0$ as $\C_h$. Let $\mathbf{p}=(x_0,y_0)$ be a point on $\C_f$.
We call $\mathbf{p}$ as an \textbf{$x$-critical point} if $f(\mathbf{p})=\frac{\partial f}{\partial
y}(\mathbf{p})=0\,$, a \textbf{singular point} if
$f(\mathbf{p})=\frac{\partial f}{\partial
y}(\mathbf{p})=\frac{\partial f}{\partial x}(\mathbf{p})=0$. $x_0$ is called an $x$-critical value of $f=0$ if $\mathbf{p}$ is an $x$-critical point of $f=0$.

Let $\mathcal{C}_{f,\,g}$ denote the algebraic
space curve defined by polynomials $f(x,y,z),
g(x,y,z)\in\mathbb{R}[x,y,z]$. We
always use $\mathcal{C}$ to replace $\mathcal {C}_{f,\,g}$ when no
ambiguity exists.

A point on an algebraic space curve $\C$ is called an \textbf{$x$-critical point} of $\C$  if its projection point is an  $x$-critical point of the projection plane curve of the algebraic space curve.

We say a graph $\mathcal{G}$ is {\bf isotopic } to an algebraic space curve $\C\subset\R^3$ if there exists a
continuous mapping $\gamma: \R^3\times[0,1] \rightarrow \R^3$ which,
for any fixed $t\in[0,1]$, is a homeomorphism $\gamma(\cdot,t)$ from
$\R^3$ to itself, and which continuously deforms $\mathcal{G}$ into $\C$:
$\gamma(\cdot,0)=id$, $\gamma(\mathcal{G},1)=\C$.

We use a set $\{\mathscr{R},\,\mathscr{B}\}$ as the
\textbf{topology (information)} of a curve $\C$ (or $\C_h$) if $\mathscr{R}$ and
$\mathscr{B}$ satisfy the following two conditions:
\begin{itemize}
  \item $\mathscr{R}$ is the set of points which are located on
$\C$ (or $\C_h$), $\mathscr{B}$ is
  the set of the connection relationship for the points in $\mathscr{R}$.
  \item If we connect the points in $\mathscr{R}$ as the connection
  relationship in $\mathscr{B}$, then the obtained graph,
  denoted as $\G$, is isotopic to $\C$ (or $\C_h$).
\end{itemize}
We list some necessary definitions and results which have been introduced in \cite{NSPC,SPC,vega} below for the consistency of the paper.
\begin{define}[Intersection of two plane curves]
Let $u,v\in \Q[x,y],\,\gcd(u,v)=1,\,w=\res_y(u,v),\res_y(u,v)$ is the resultant of polynomials $u,\,v$ \wrt $y$. We say $u,v$ are in a {\bf generic
position} w.r.t. $y$ if the following two conditions are satisfied, where $lc(T,y)$ is the leading coefficient of $T$ \wrt $y$. \\
\indent 1) $\gcd(\mbox{lc}(u,y),\mbox{lc}(v,y))=1$.\\
\indent 2) $\forall\alpha\in \mathbb{V}_{\CC}(w)$,
$u(\alpha,y),v(\alpha,y)$ have only one common zero in $\CC$.
\end{define}
\begin{define}
Let $h$ be a squarefree polynomial in $\Q[x, y]$. The real algebraic plane curve defined by $h$,\,say $\C_h$, is in generic position \wrt $y$ if the following two conditions are verified:\\
\indent 1) The leading coefficient of $h$ \wrt $y$ is a constant.\\
\indent 2) $\forall\alpha\in \R$,
$h(\alpha,y),h_y(\alpha,y)$ have at most one common distinct zero in $\CC$.
\end{define}
According to the above definitions, the following corollary is immediate.
\begin{corollary}
Let $f\in \Q[x,y]$ be a square free polynomial, $lc(h,y)$ is a constant, then the plane curve $\C_h$ is in a generic position \wrt $y$ if and only if the intersections of $h,h_y$ is in a generic position.
\end{corollary}
For a plane curve $\C_h$, we can get a set $S\subset\R$ s.t. the plane curve $\C_{\bar{h}}$ is in a generic position for $\forall s\in S$ by resultant computation, where $\bar{h}=h(x+sy,y)$. It can be proceed as below: $\bar{h}=\mbox{sqrfree}(\bar{h})$,
let $q(x,s)=\mbox{sqrfree}(\res_y(\bar{h},\bar{h}_y))$,
$Q(s)=\mbox{sqrfree}(\res_x(q,q_x))$, then $S=\R\setminus\V_{\R}(Q)$, see \cite{DBouziane,ElKahoui97}.

The method above can be generalized as the following algorithm. We can simply call the algorithm ${\bf PCGP}(s)$.

 {\it PCGP($s$)}: Given a plane curve with a parameter $s$, say $h(x,y,s)=0$, our aim is to find a proper $s_0\in \Q$ s.t. the plane curve $h(x,y,s_0)=0$ is in a generic position.

We will use this technique to compute a similar set $S\subset \Q$ s.t. the sheared algebraic space curve is in a generic position under the shearing related to any number in $S$ later. First, we give some relevant definitions.
\begin{define}[\cite{NSPC,SPC}]
Let $f,g\in\Q[x,y,z]$ such that $\gcd(f,g)=1$, $h=\mbox{sqrfree}(\res_z(f,g))$, we say $f,g$ are in a {\bf weak generic position} \wrt $z$ if \\
\indent 1) $\gcd(\mbox{lc}(f,z),\mbox{lc}(g,z))=1$.\\
\indent 2) There are only a finite number of $(\alpha,\,\beta)\in\mathbb{V}_{\CC}(h)\subset\CC^2$
such that $f(\alpha,\beta,z),g(\alpha,\beta,z)$ have more than one distinct
common zeros in $\CC$.
\end{define}

The definition of generic position for an algebraic space curve here is weaker than that appears in \cite{dia}, where not only the plane projection curve should be in a generic position, but the apparent singular point at each fiber in the plane curve should be a node. It is close to the definition {\it Pseudo-generic position} in \cite{dia}. In the following, we will provide a simple method to check weak generic position for an algebraic space curve.
\begin{theorem}[\cite{NSPC}]\label{GPtheorem}
If $gcd(lc(f,z),lc(g,z))=u$ be a constant, then $f(x,y,z),g(x,y,z)$ is in a weak generic
position \wrt $z$ if and only if $\exists\,\alpha\in\CC$, such
that $f(\alpha,y,z),\,g(\alpha,y,z)$ is in a generic position
w.r.t.~$z$.
\end{theorem}
According to the above theorem, we can transpose the problem of weak generic position checking of an algebraic space curve into the generic position checking of the intersection of two plane algebraic curves.

According to Theorem \ref{GPtheorem}, it is equivalent to choose an $\alpha$ and $S_1$ s.t. $f(\alpha,y+sz,z)\wedge g(\alpha,y+sz,z)$ is in a generic position \wrt $z,\forall s\in S_1$. It can be proceed as follow:\\
Let $q_1(y,s)=\mbox{sqrfree}(\res_z(F(\alpha,y+sz,z),G(\alpha,y+sz,z)))$, and compute $Q_1(s)=\mbox{sqrfree}(\res_y(q_1,\frac{\partial q_1}{\partial y}))$. Then the set $S_1$ can be chosen as the complementary set of $\V_{\R}(Q_1(s))$ in $\R$. In the practical computation, the default value of $\alpha$ is $0$.

\section{Certifying Generic Position }\label{GPchecking}
In this section, we will present an algorithm to compute a deterministic generic position for an algebraic space curve.

\begin{define}\label{GenericPosition}
Let $f,\,g\in\Q[x,\,y,\,z]$ be squarefree polynomials. The algebraic space curve defined by $f,\,g$,\,denoted as $\C$, is called in a {\bf generic position} w.r.t. $z$
if \\
\indent 1) $f,g$ are in a {\bf weak generic position} w.r.t. $z$\\
\indent 2) The projected plane curve $\C_h$ is in a generic position \wrt $y$, where $h=\mbox{sqrfree}(\res_z(f,g))$.
\end{define}
For an algebraic space curve $\C$ in a generic position, two $x$-critical points of $\C$ may correspond to the same projection $x$-critical point of $\C_h$. We prefer to remove this case. So we give the following definition.
\begin{define}\label{def-StrongGP}
Let $f,\,g\in\Q[x,\,y,\,z]$ be squarefree polynomials. The algebraic space curve defined by $f,\,g$,\,denoted as $\C$, is called in a {\bf  strong generic position} w.r.t. $z$
if \\
\indent 1) $f,g$ are in a {\bf generic position} w.r.t. $z$\\
\indent 2) For each $x$-critical point $p$ of $\C_h$, there is at most one $x$-critical points of $\C$ whose projection is $p$.
\end{define}

We give an algorithm below to compute a deterministic strong generic position for an algebraic space curve.
\begin{algorithm}
\caption{Find-strongGP}\label{Find-strongGP}
\KwIn{$f(x,y,z),g(x,y,z)\in\Q[x,y,z]$. }
\KwOut{$F\wedge G$ (isotopic to $f\wedge g$) is in a strong generic position.}
If $\deg(f)\neq \deg(f,z)$ and $\deg(g)\neq \deg(g,z)$,
$f:=f(x+z,y+z,z),\,g:=g(x+z,y+z,z)$\;
Let $h_1(x,y,s_1):=\Res_z(f(x+s_1y,y,z), g(x+s_1y,y,z))$.
Call ${\it PCGP}(s_1)$ (see Section 2) to find $s_1\in\Q$ s.t. the projected curve of $f(x+s_1y,y,z)\wedge g(x+s_1y,y,z)$ is in a generic position \wrt $y$. Still denote the obtain new algebraic space curve as $f\wedge g$\;

Let $h_2(x,y,s_2):=\Res_z(f(x+s_2z,y,z), g(x+s_2z,y,z))$.
Call ${\it PCGP}(s_2)$ to find $s_2\in\Q$ s.t. the $x$-coordinates of the $x$-critical points of the algebraic space curve $f(x+s_2z,y,z)\wedge g(x+s_2z,y,z)$ are distinct. Still denote the obtain algebraic space curve as $f\wedge g$\;

Let $h_3(x,y,s_3):=\Res_z(f(x, y+s_3z,z), g(x, y+s_3z,z))$. For $\forall \alpha\in\Q$, find $s_3\in\Q$ s.t. \\
\begin{enumerate}
\item the intersection of $f(\alpha,y+s_3z,z),g(\alpha,y+s_3z,z))$ is in a generic position \wrt $z$.\\
\item Call ${\it PCGP}(s_3)$ to find $s_3\in\Q$ s.t. the plane projected curve of $f(x, y+s_3\,z,z)\wedge g(x, y+s_3\,z,z)$ is in a generic position \wrt $y$.
\end{enumerate}
Denote the space curve as $F\wedge G$\;
\textbf{Return} $F,G$.
\end{algorithm}
\begin{theorem}
Algorithm \ref{Find-strongGP} is correct and it terminates in finite steps.
\end{theorem}
\proof The termination of the Algorithm \ref{Find-strongGP} is clear. We need only to prove the correctness of the algorithm. The first linear coordinate transformation makes the gcd of the leading coefficients \wrt $z$ of the defining polynomials to be a constant.  The second coordinate transformation  of Algorithm \ref{Find-strongGP} puts the projected plane curve in a generic position \wrt $y$; After the transformation, the $x$-critical points of the new space curve are with different $x$-coordinates except for the case their $(x,y)$ coordinate pair are the same. The third coordinate transformation makes the $x$-coordinates of all the $x$-critical points of the sheared  algebraic space curve to be distinct. Note that the projection curve of the space curve after the second or third coordinate transformations both are in a generic position. The last coordinate transformation ensures that the space curve is in a weak generic position \wrt $z$ and the projection curve is in a generic position. According to the definition \ref{def-StrongGP}, we know the output space curve $F\wedge G$ is in a strong generic position \wrt $z$.
\qed


As we know, this is the first time to find a deterministic (strong) generic position for an algebraic space curve, not just checking the generic position as presented in \cite{JGA05,SPC,dia,ElKahoui08}.

\section{Outline of the Algorithm}
In this section, we consider the topology computation for an algebraic space
curve $f(x,y,z)\wedge g(x,y,z), f,g\in\Z[x,y,z]$. The main steps of the algorithm are presented in \cite{NSPC}. We modify the lifting step in the algorithm in \cite{NSPC} and add a preparatory step. Thus there is no assumption for the input algebraic space curve.

%

\noindent{\textbf{Overview of the algorithm.}}
\begin{enumerate}
\item Certified generic position: Call Algorithm \ref{Find-strongGP}. We denote $\C=\{(x,y,z)\in\R^3|F(x,y,z)=G(x,y,z)=0\}$, where $F$ and $G$ are the output polynomials.
\item Projection: Compute the topology of the projection plane curve $\C_h$.
\item Lifting: Get the space point candidates which may contain points of $\C$.
\item Compute $s$: First, we compute a set $S'$ such that every two sheared space point candidates (obtained from step $3$) will not overlap  after linear coordinate transformation and projection for any $ s\in S'$. Then we compute $S''$ s.t. $\forall s\in S''$ the sheared algebraic space curve $F(x,y+sz,z)\wedge G(x,y+sz,z)$ is in a weak generic position \wrt $z$. Let $S=S'\cap S''$, and we choose a rational number $s\in S$.
\item Compute the topology of $\C_{\bar{h}}$: Let $\overline{\mathcal {C}}=\big\{(x,y,z)\in\mathbb{R}^3\mid
    \overline{F}=F(x,y+sz,z)=0, \overline{G}=G(x,y+sz,z)=0\big\}$
  and  project the space curve $\overline{\C}$ into the
   $xy$-plane and compute the topology of the projected curve
   $\mathcal {C}_{\bar{h}}$, where $\bar{h}=\mbox{sqrfree}(\Res_z(\overline{F},\overline{G}))$.
\item Computing the refined topologies and certifying: For the $x$-critical values of $\C_h$ which does not appear in the $x$-critical values of $\C_{\bar{h}}$, we use the corresponding vertical lines to intersect $\C_{\bar{h}}$ and add the points in the topology information of $\C_{\bar{h}}$. We do the similar operation on $\C_h$.
Then we certify the space point candidates by comparing the
topology information of $\mathcal {C}_{h}$ and $\mathcal
{C}_{\bar{h}}$.
\item Connection: Connect the space points 
by line segments in an appropriate manner (by comparing the connection of the points of $\C_h$ and $\C_{\bar{h}}$). Hence we get the topology of the algebraic space curve.
\end{enumerate}

\noindent{\bf Remark:} In Step 1, when we call Algorithm 1, the last step in Step 4.2 can be removed since $F\wedge G$ in Step 1 can be not in a weak generic position.

\subsection{Certified generic position}
 Please see Algorithm \ref{Find-strongGP} and Remark above.
%
\subsection{Computing the topology of $\C_h$~($\C_{\bar{h}}$)}\label{section-computetop}
In this step, we will compute the topology of algebraic plane curves $\C_h,\,\C_{\bar{h}}$. We consider only $\C_h$ for example. First, project the space curve $\mathcal {C}$ onto
the $xy$-plane to get a plane curve $\mathcal {C}_h$, where
$h=\mbox{sqrfree}(\Res_z(F,G))$.
Then the topology information $\{\mathscr{R}_h,\mathscr{B}_h\}$ of  $\mathcal {C}_h$ can be easily computed. Actually, there are many papers \cite{amj,Berberich11,top-cur,eigenwillig08,eigenwillig,hong,Kerber,vega,Seidel2005} dealing with the topology computation of a plane curve. We adopt
the methods presented in \cite{Kerber} in this paper. In the process of
topology computation of $\C_h$,
$p(x)=\mbox{sqrfree}(\Res_y(h,h_y))$ is also obtained.
Assuming $\mathscr{R}_h=\big\{(\alpha_i,\beta_{i,j})|
i=1,\ldots,l_0,\, j=1,\ldots,l_i, \alpha_k<\alpha_{k+1}(1\le k<l_0) ,\, \beta_{i,k}<\beta_{i,k+1}(1\le k< l_i)\big\},$ $l_0$ is the number of the
real roots for $p(x),\,l_i$ is the number of the real roots for
$h(\alpha_i,\,y),$ where $p(\alpha_i)=0$.
We use isolating interval $I_i\times J_{i,j}$ to represent the point $(\alpha_i,\beta_{i,j}),\forall i\in[l_0],\forall j\in[l_i]$.

\subsection{Lifting}\label{lifting}
After we have computed the topology of the plane curve $\C_h$, a
main difficulty for computing the topology of $\mathcal {C}$ is to
obtain the space points lifted from the points in $\mathscr{R}_h$. It can be divided into two cases, we will discuss it below.\\
(i) The lifting of the $x$-critical points of $\C_h$. For $\forall(\alpha_i,\beta_{i,j})\in\V_{\R}(h(x,y),h_y(x,y))$, assume its isolating box is $I_i\times J_{i,j}$. We isolate the real roots of interval polynomial equations $f(I_i,J_{i,j},z)=0$ and $g(I_i,J_{i,j},z)=0$ respectively, whose isolating intervals are
$\tilde{K}_{i,j,k},k\in[l_{i,j,1}]$ and ${K'}_{i,j,k},\,k\in[l_{i,j,2}]$. As to the method of isolating the real roots of interval polynomials, please refer to \cite{GPC}. Assume
$\big\{\cup_{k=1}^{l_{i,j,1}}\tilde{K}_{i,j,k}\big\}
\cap \big\{\cup_{k=1}^{l_{i,j,2}}{K'}_{i,j,k}\big\}=
\big\{\cup_{k=1}^{l_{i,j,3}}{K}_{i,j,k}\big\}$.\\
(ii)The lifting of the regular points on $\C_h$. As to the lifting of the regular points at the $\alpha_i$ fiber, we choose a rational number $t_i\in I_i$, and solve the bivariate system $\{F(t_i,y,z)=G(t_i,y,z)=0\}$. This system can be efficiently solved by the methods presented in \cite{Bouzidi,Emeliyanenko,GPC}, the bit complexity of the first two is $\O(d^7(d+\tau))$ while the third one is $\O(d^9(d+\tau))$~(for definitions of $d$ and $\tau$, see details in section \ref{Complexityanalysis}).\\
After the above treatment, we
obtain all candidate boxes of the possible space points and denote
these boxes collection as $SBS=\{\mathbb{I}_{i,j,k}=I_i\times J_{i,j}\times
K_{i,j,k},i\in[l_0],
j\in [l_i],k\in[l_{i,j}]\}.$ Then what we want to do is
to certify which boxes of $SBS$ really contain space
points or not. This will be checked later. In fact, the candidates obtained in (ii) are already certified.

\subsection{Computing s}
Before checking the space point candidates, we will compute an
interval set $S\subset\R$ such that for $s\in S$ it satisfies the following two conditions:\\
(i)The projections of the sheared uncertified elements in $SBS$ on the $xy$-plane can be easily certified by the new projection curve of the sheared space curve. \\
(ii)The algebraic space curve defined by the polynomials $F(x,y+sz,z)$ and
$G(x,y+sz,z)$ in a weak generic position \wrt $z$.\\
The second goal can be easily achieved using the technique of Section \ref{GPchecking}, and we denote the obtained interval as $S''$.

As to the first one, a simple way is to ensure that the projections of all the sheared elements in $SBS$ on the $xy$-plane have no intersections each other. But we can simplify the computation as below:\\
For $\forall i\in[l_0]$, we define a pair set $P_i$, accurately
$P_i=\{(j,\,k)| j\in[l_{i}],k\in[l_{i,j}]\}$. Since $\C_h$ is in a generic position, there is only one $x$-critical point of $\C_h$ on the fibre related to $i$. We assume that the space point candidates related to the $x$-critical point of $\C_h$ are $P_i^1=\{(j^0,k) \in P_i, k\in [l_{i,j^0}]\}$. They may be not certified (Some of them may be certified, see \cite{GPC}). But the space point candidates in $P_i\setminus P_i^1$ are certified.

Let
$S_{i}=\big\{s\in\R|
(J_{i,j_0}+s\,K_{i,j_0,k_0})
\cap(J_{i,j_1}+s\,K_{i,j_1,k_1})=\emptyset,
\forall (j_0,k_0)\in P_i^1,(j_1,k_1)\in P_i\setminus P_i^1\big\}\cap\{s\in\R|(J_{i,j^0}+s\,K_{i,j^0,k_0})
\cap(J_{i,j^0}+s\,K_{i,j^0,k_1})=\emptyset,
\forall (j^0,k_0),(j^0,k_1)\in P_i^1, k_0\neq k_1\}$, and
$S'=\bigcap_{i=1}^{l_0}S_i$. Note that there is at most one $x$-critical point of $\bar{\C}$ related to the $x$-critical point of $\C_{\bar{h}}$. We can always find a nonempty $S'$.
We denote $S$ as the intersection of $S'$ and $S''$.

Now we choose a simple rational number $s\in S$ and define two maps
$\phi_s$ and $\pi$.
$$\begin{array}{cccc}
    \phi_s: & \R^3 & \longrightarrow & \R^3 \\
    {} & (x,y,z) & \longrightarrow & (x,y+sz,z)\\
    \pi:& \R^3 &\longrightarrow & \R^2 \\
    {} & (x,y,z)&\longrightarrow & (x,y)
  \end{array}
$$
Let
$\overline{F}(x,y,z)=\phi_s(F)=F(x,y+sz,z),\,\overline{G}(x,y,z)=\phi_s(G)=G(x,y+sz,z),\,$
and $\overline{\mathcal {C}}$ is used to denote the sheared space curve
$\overline{F}\wedge \overline{G}$ in the rest parts of this paper.
\begin{equation*}\label{sheer-curve}
    \bar{h}(x,y)=\mbox{sqrfree}(\Res_z(\overline{F}(x,y,z),\,\overline{G}(x,y,z)))
\end{equation*}
Then, we compute the topology of the plane curve
$\C_{\bar{h}}$ as Section \ref{section-computetop}. In the nest subsection, we will use the points on $\C_{\bar{h}}$ to certify the space point candidates in $SBS$.
\subsection{Computing the two refined topologies and Certifying}

Now, we will compute the refined topology
information for the plane curves of $\C_h$ and $\C_{\bar{h}}$ such
that the numbers of $x$-fibers in their refined topology information
are equal.


With the above preparation, we can show how to certify all the
candidate boxes in $SBS$. The main idea for certification is comparing the points on the topology of $\C_h$ and $\C_{\bar{h}}$. We just give a simple introduction for the certification of the elements in $SBS$, for more details, please refer to \cite{NSPC}.

For each space point candidate $\mathbb{I}_{i,j,k}=I_i\times J_{i,j}\times
K_{i,j,k}\in
SBS,$ we assume that its image is
$I_i\times J_{i,j,k}\times
K_{i,j,k}$
under $\phi_{-s},$ where
$J_{i,j,k}=J_{i,j}-sK_{i,j,k}$.
Then we project its image onto the $xy$-plane which leads to a
box $I_i\times J_{i,j,k}.$
Meanwhile, for the
same $i$ we can get a disjoint interval set from the topology
information of the curve $\C_{\bar{h}}$, that is, the isolating intervals of the real roots of $\bar{h}(\alpha_i,\,y)$. Let this interval set be
$\bar{J}_i=\big\{\bar{J}_{i,j}
\big|j\in[\bar{l}_i]\big\}$, where $\bar{l}_i$ is the number of the distinct real roots of $\bar{h}(\alpha_i,y)$. The certification of space point candidates are achieved by comparing each $J_{i,j,k}$ with the interval set $\bar{J}_i$. See \cite{NSPC,GPC} for more details.
%
%

%
After this operation, the superfluous candidate boxes are deleted and the left
space point candidates have unique correspondence with the intervals in
$\bar{J}_i$. Then the left thing is to
determine the connection relationship of space points between
each two adjacent $x$-fibers, and it will be discussed in the next subsection.

\subsection{Connecting and space curve recovery}
\label{space-curv-topo-rec}
Let us recall the information of what we have obtained. That is, the
topology information of two plane curves $\mathcal {C}_{h}$ and
$\mathcal {C}_{\bar{h}}$, the space points lifted from the plane points on $\C_h$ and the correspondence relationship between these $3$-d points and the $2$-points on
$\mathcal {C}_{\bar{h}}$. As a matter of fact, this
information is enough to recover the topology of $\mathcal {C}$
accurately. 

Recovering the topology of the space curve means that we should
find the connection relationship between the space points on each two
adjacent parallel planes with the form $x-\alpha=0$. Both $\C_h$ and $\C_{\bar{h}}$ are in a generic position \wrt $y$, $\overline{\C}$ is in a generic position \wrt $z$, and the certified space points have a one to one correspondence with the points on $\C_{\bar{h}}$. Thus the connection of the space points can be obtained by the connection of $\C_{\bar{h}}$ and $\C_h$. If necessary, we can add a vertical plane between two adjacent $x$-fibers to get the connection. In a word, we can get the connection relationship of the space points. For more details, please refer to \cite{NSPC,SPC}.

Thus, we get the topology of $\mathcal {C}$.


\section{Complexity analysis}\label{Complexityanalysis}
In this section, we will analyze the complexity for computing the
topology of an algebraic space curve. In this paper, all the complexity means the bit complexity, and the $\tilde{\mathcal {O}}$-notation means
that we ignore logarithmic factors. The complexity for computing the
topology of a plane curve has been studied by several of previous
algorithms. See
\cite{D-arnon,vega-and-Kah,S-Basu,eq2-dio,Kerber},
and the best complexity is $\tilde{\mathcal {O}}(N^{10})$ which is
given in \cite{Kerber}, where
$N=\max\{d,\,\tau\}$, and $d$ is the degree bound for the input polynomial while  $\tau$ is the bitsize bound for the coefficients of the polynomial. In the following part, we will show
that the bit complexity of our algorithm is bounded by
$\tilde{\mathcal {O}}(d^{19}(d+\tau))$ but without generic position
assumption. This is the best complexity that improves the previous best known complexity bounds by a factor of $d^2$. First, we will give some notations and results for complexity analysis.
\subsection{Basic results for complexity}
For a univariate polynomial $f=\sum_{i=0}^{n}a_ix^i\in\Z[x]$ with roots
$z_1,\ldots,z_n\in\mathbb{C}$ and $a_n\neq 0$, $n=\deg (f)$ denotes
its degree. The separation bound of $f$ is defined as
$\mbox{sep}(f)=\min_{z_i\neq z_j}\mid z_i-z_j\mid$. We take the conventions in \cite{Kerber} that an integer
polynomial is called of magnitude $(d,\tau)$ if its total degree
is bounded by $d$, and each integer coefficient is bounded by
$2^{\tau}$ in its absolute value. For simplicity, we mostly ignore
the logarithmic factors in $d$ and $\tau$ in the complexity bounds.
$\tilde{\mathcal {O}}$ indicates that we omit logarithmic factors.
For $a \in \mathbb{Q}$, $\mathcal {L}(a)$ is the maximum bitsize of
the numerator and the denominator. First, we list the complexity of
several basic operations on univariate and bivariate polynomials.
\begin{define}
For an interval $I=[a,b]$, we define its bitsize to be the maximal
bitsize of the endpoints of the interval, that is,
$\LL(I)=\max{(\LL(a),\LL(b))}$; Moreover, if a real number $\xi$ is
represented by an interval $I=[a,b]$, then the bitsize of $\xi$ is
defined to be $\LL(I)$.
\end{define}


\begin{lemma}[see\cite{Elias2013,Sagraloff1}]\label{lemstrongiso}
For a square-free polynomial $f$ of degree $n$ with integer
coefficients of modulus less than $2^\tau,$ we can compute isolating
intervals for all real roots of $f$ (for a single real root of $f$) of width less than $2^{-L}$
using no more than $\O(n^3\tau+n^2L)$ ($\O(n^2\tau+n\,L)$)bit operations.
\end{lemma}

\begin{lemma}[Kerber 2012 \cite{Kerber}]\label{sqrfree}
Let $g \in \Z[x]$ be a polynomial with degree $d$
and bitsize $\lambda$. Its square-free part $g^{*}$ can be computed
in $\O(d^2\lambda)$, and it has degree at most $d$. Its bitsize of
each coefficient of $g^{*}$ is bounded by $\O(d+\lambda).$
\end{lemma}

\begin{lemma}[\cite{S-Basu}]\label{prodp}
If both the magnitude of the univariate polynomials $u(x)$ and
$u_1(x)$ are $(d,\,\tau)$, then the magnitude of the  product
polynomial $u(x)\cdot u_1(x)$ is $(d,\,\tau)$.
\end{lemma}

\begin{lemma}[univariate polynomial evaluation
\cite{S-Basu}]\label{univar-evaluation} If $g\in\Z[x]$ is of
magnitude $(d,\,\tau)$, and a rational number $a$ with bitsize
$\sigma$, then evaluating $g(a)$ has a complexity of
$\O(d(d\tau+\sigma))$, and the bitsize of $g(a)$ is $\mathcal
{O}(d\sigma+\tau)$.
\end{lemma}
According to Lemma~\ref{univar-evaluation}, we have the following
corollary:
\begin{corollary}[multivariate~polynomial~evaluation]\label{multivar-evaluation}
Let $g\in\Z[x_1,\ldots,x_m]$ is of
magnitude $(d,\,\tau)$, and $m\geq 1$ rational numbers
$a_1,\ldots,a_m$ all with bitsize $\sigma$, then evaluating
$g(a_1,\ldots,a_m)$ has a complexity of $\O(d^m(d\sigma+\tau))$, and
the bitsize of $g(a_1,\ldots,a_m)$ is $\mathcal {O}(d\sigma+\tau)$.
\end{corollary}
\proof Its correctness follows easily that there are $\mathcal
{O}(d^m)$ arithmetic operations and the bitsize of each number is
bounded by $\mathcal {O}(d\sigma+\tau)$. Thus the bit complexity is
$\O(d^m(d\sigma+\tau))$. \qed

\begin{corollary}\label{m-u-evaluation}
Let $g\in\Z[x_1,\ldots,x_m]$ be a multivariate polynomial with
degree $d$ and coefficients bounded by $2^{\tau}$, and
$m-1,(m\geq2)$ rational numbers $a_1,\ldots,a_{m-1}$ all with
bitsize $\sigma$, then evaluating $g(a_1,\ldots,a_{m-1},x_m)$ has a
bit complexity of $\O(d^m(d\sigma+\tau))$, and the resulting
univariate polynomial is of magnitude $(d,\,d\sigma+\tau)$.
\end{corollary}

\begin{lemma}\label{lem-sep}
\cite{S-Basu,mignotte,c-yap}\,Let $f(x)$ be a polynomial in $\Z[x]$
and $\deg_{x}(f)\leq d$, $\LL(f)\leq \tau$. Then the separation
bound of $f$ is
\begin{equation*}
    \mbox{sep}(f)\geq d^{-\frac{d+2}{2}}(d+1)^{\frac{1-d}{2}}2^{\tau
(1-d)},
\end{equation*}
thus $\log(\frac{1}{\mbox{sep}(f)})=\tilde{\mathcal {O}}(d\tau)$.
The latter provides a bound on the bitsize of the endpoints of the
isolating intervals.
\end{lemma}

\begin{lemma}\label{lem-Dim2}
\cite{eq2-dio}\,Let $f, g \in (\Z[y_1,\ldots , y_k])[x]$ with
$\deg_x(f)=p\geq q=\deg_x(g)$, $\deg_{y_i}(f)\leq p$ and
$\deg_{y_i}(g) \leq q$, $\mathcal {L}(f)=\tau\geq\sigma=\mathcal
{L}(g)$. We can compute $ \res_x(f, g)$ in $\tilde{\mathcal
{O}} (q(p+q)^{k+1}p^k\tau)$. And $\deg_{y_i}(\res_x(f,g))\leq 2pq$,
and the bitsize of coefficients for the resultant is
$\tilde{\mathcal {O}}(p\sigma+q\tau).$
\end{lemma}
\begin{theorem}[\cite{Bouzidi,Emeliyanenko}]\label{bivariatesolving}
Let $f_1(x,y),\,f_2(x,y)\in\Z[x,y]$ are of magnitude $(d,\tau)$, and they have only trivial common factor in $\CC[x,y]$. Then their real roots can be computed with $\O(d^7(d+\tau))$.
\end{theorem}
We remark that in \cite{GPC}, we also present an efficient algorithm for solving zero-dimensional polynomial systems. The bit complexity for bivariate case is $\O(d^9(d+\tau))$. For the bit complexity of an algebraic plane curve, we have the following:
\begin{theorem}[\cite{Kerber}]\label{plane_top}
If $f(x,\,y)\in\Q[x,y]$ with magnitude of $(d,\tau)$, then we can
compute the topology of plane curve $\C_f$ in $\O(d^{9}(d+\tau))$ bit
operations.
\end{theorem}
\subsection{Main results for the complexity of the algorithm}
In this subsection, we will give the complexity analysis for the algorithm step by step.

%
\begin{theorem}\label{thm-complexity-StrongGP}
Let $f,g\in \Z[x,y,z]$ with magnitude of $(d,\tau)$. The bit complexity of Algorithm \ref{Find-strongGP} is bounded by $\O(N^{20})$, where $N=\max\{d,\tau\}$. That is, we can find a deterministic strong generic position for an algebraic space curve in $\O(N^{20})$ bit complexity.
\end{theorem}
\proof In Algorithm \ref{Find-strongGP}, the bit complexity of Step 1 and Step 4.1 can be ignored compared to other steps. It is not difficult to find that the bit complexity of Step 2, Step 3 and Step 4.2 are similar. So we just consider the complexity of one coordinate transformation, say $\{x,y,z\}\rightarrow\{x,y+sz,z\}$. The sheared polynomials $f(x,y+sz,z), g(x,y+sz,z)$ are with magnitude $(2d,\O(\tau))$. Then we compute $h(x,y,s)=\mbox{sqrfree}(\res_z(f(x,y+sz,z),g(x,y+sz,z)))$. This computation needs $\O(d(2d+d)^{2*3+1}\tau)=\O(d^8\tau)$ bit complexity by Lemma \ref{lem-Dim2}, and $h(x,y,s)$ is of magnitude $(d^2,d\tau)$ when ignoring the logarithmic factors. According to the plane generic position checking in \cite{GPC,eq2-dio}, we know it costs $$\O((d^2)^{10}+(d^2)^{9}d\tau)=\O(d^{20}+d^{19}\tau)$$ bit complexity, which is bounded by $\O(N^{20})$, where $N=\max\{d,\tau\}$.
So we prove the theorem.
\qed

It is obvious that the bitsize of $s$ is $\OO(\log d)$. Thus the magnitude of the sheared polynomials are of $(2d,\OO(d\log d+\tau))=(\OO(d),\O(d+\tau))$.
\begin{lemma}\label{liftingThm}
The complexity of the lifting process is bounded by $\O(d^{19}(d+\tau))$, and the bitsize for the endpoints of the space point candidates is bounded by $\OO(d^{11}\tau)$ .
\end{lemma}
\proof We consider the $x$-critical value $\alpha_i\in I_i=[a_i,\,b_i]$ as an illustration. As discussed in Subsection \ref{lifting}, we divide it into two steps.
First, we consider the
isolating boxes for the real root of the system
$\Sigma=\big\{h(x,\,y)=h_y(x,\,y)=0\big\}$, solving this system needs $$\O((d^2)^9(d^2+d\tau))=\O(d^{19}(d+\tau))$$
bit complexity and the bitsize of the
endpoints of those isolating boxes can be controlled by $\mathcal
{O}(d^7\tau)$ using the results in \cite{GPC} since the univariate polynomials are with magnitude of $(d^4,d^3\tau)$. Then these boxes
are substituted into the polynomials $F(x,y,z)$ and $G(x,y,z)$, and
the resulting interval polynomials $F(I_i,\,J_{i,j},z)$ and
$G(I_i,\,J_{i,j},z)$ are with degree $d$ and coefficients bounded by
$\O(d^8\tau)$. Thus the bitsize of the $z$ coordinate is bounded by
$\O(d^9\tau)$, and the evaluation of $F(I_i,J_{i,j},z)$ yields
$\O(d^3(d^7\tau+\tau))=\O(d^{10}\tau)$ bit complexity according to
the Corollary \ref{m-u-evaluation}. By Lemma \ref{lemstrongiso}, we know the complexity of isolating the interval polynomials is $\O(4d^3\cdot d^8\tau)=\O(d^{11}\tau)$. Note that there are only one $x$-critical point in each $x$ fiber. So we get the space points candidates
lifted from $x$-critical value $\alpha_i$ of the plane curve $\C_h$.

In order to get the space candidates lifted from the regular points of the
topology information of $\C_h$, we do the following consideration.
First we choose a rational value $t_i\in I_i$, and solve the
bivariate system $\{F(t_i,y,z)=G(t_i,y,z)=0\}$. Both the magnitudes of the
polynomials $F(t_i,y,z)$ and $G(t_i,y,z)$ are of $(d,\,d^8\tau)$,
solving this system requires
$$\O(d^{7}(d+d^8\tau))=\O(d^{15}\tau)$$
bit complexity according to Theorem \ref{bivariatesolving}, and the bitsize of the obtained isolating boxes is
bounded by $\O(d^{11}\tau)$.
By comparing the solutions of
$\{F(t_i,y,z)=G(t_i,y,z)=0\}$ and the branch numbers at the
$\alpha_i$ fiber of $\C_h$, we can easily get the space candidates
lifted from the regular points, and the bitsizes of these boxes are
bounded by $\OO(d^{11}\tau)$.

So far, we get all the space
candidate at the $\alpha_i$ fiber and the total bit complexity is
bounded by $\O(d^{15}\tau+d^8)$. There are $\OO(d^4)$ fibers.
Thus, the total bit complexity to get all the space point candidates is
$$\OO(d^4)\cdot\O(d^{15}\tau)=\O(d^{19}\tau),$$
and the bitsizes of the
candidates are $\OO(d^{11}\tau)$, this completes the lemma.
\qed

Now we analyze the complexity of computing $S$.
\begin{lemma}\label{computingsi}
The complexity of computing $S$ is bounded by $\O(N^{20})$, where $N=\max\{d,\tau\}$.
\end{lemma}
\proof According to the Theorem \ref{thm-complexity-StrongGP}, we know the complexity of computing $S''$ is $\O(d^{19}(d+\tau))$. Now we consider only the process of computing  $S'$. In each $x=\alpha_i$ fiber, there are  $\OO(d^2)$
candidates at most. Then for every two candidates, we just need $\OO(1)$ multiplications. So the complexity is
$\OO(1)\O(d^{11}\tau)=\O(d^{11}\tau)$ if we use fast Fourier transform since the bitsize of the endpoint is bounded by $\O(d^{11}\tau)$. When any two space point candidates on a fiber $x=\alpha$ do a similar operation, there are at most $C_{\OO(d^2)}^2=\OO(d^4)$ combinations, which implies  the bit complexity for computing $S_i$ is bounded by $\OO(d^4)
\O(d^{11}\tau)=\tilde{\mathcal {O}}(d^{15}\tau)$.

There are $\OO(d^4)$ fibers at most, thus the total complexity for computing $S'$ is bounded by
$$\OO(d^4)\O(d^{15}\tau)=\O(d^{19}\tau).$$
So the total complexity is dominated by $\O(d^{19}(d+\tau))$, which is as desired.
\qed

As  to the connection process, we have the following lemma.
\begin{lemma}\label{Connectionthm}
The bit complexity for the connection step is bounded by $\O(d^{19}\tau)$.
\end{lemma}
\proof In the process of the
connection determination, for the most cases, we can connect correctly easily by comparing the connection of the points in $\C_h$ and $\C_{\bar{h}}$. As to the difficult case, we take the strategy which has been used in \cite{SPC,dia,Diatta}. That is, adding an auxiliary number
between each two $x$-critical values to certify the connection.  It can be achieved by solving bivariate polynomials system $\{F(t_i,y,z)=G(t_i,y,z)=0\}$, where $\alpha_i<t_i<\alpha_{i+1}$ and the bitsize of $t_i$ is $\OO(d^7\tau)$, both $F(t_i,y,z)$ and $G(t_i,y,z)$ are with magnitudes of $(d,d^8\tau)$. So It needs $$\O(d^7(d+d^8\tau))=\O(d^{15}\tau)$$
bit complexity by Theorem \ref{bivariatesolving}.  There are $\OO(d^4)$ fibers, thus the total complexity for connection process is bounded by $\OO(d^4)\O(d^{15}\tau)=\O(d^{19}\tau)$.
\qed
\begin{theorem}
Let $f,g\in \Z[x,y,z]$ be with magnitude of $(d,\tau)$,
then we can compute the topology of the algebraic space curve defined by the
polynomials $f(x,y,z)$ and $g(x,y,z)$ in $\tilde{\mathcal
{O}}(N^{20})$ bit operations, where $N=\max\{d,\tau\}$.
\end{theorem}
\proof First we do coordinate transformation to $f\wedge g$ s.t. the sheared space curve is in a generic position \wrt $z$, and its complexity is bounded by $\O(N^{20})$ according to Theorem \ref{thm-complexity-StrongGP}. Moreover, the magnitude of $F$ and $G$ are also $(d,\tau)$ if we ignore the logarithmic factors.

Let $h=\mbox{sqrfree}(\res_z(F,G))$, and we compute $h$
using $\tilde{\mathcal {O}}(d(d+d)^{2+1}d^2\tau)=\tilde{\mathcal
{O}}(d^6\tau)$ bit operations, and the magnitude of $h$ is
$(d^2,2d\tau)$ according to Lemma \ref{lem-Dim2}. Then we compute
the topology of the plane curve $\C_h$ in $\tilde{\mathcal
{O}}((d^2)^9(d^2+2d\tau))=\tilde{\mathcal {O}}(d^{19}(d+\tau))$ bit
operations due to Theorem \ref{plane_top}.  Assume the points in
$\mathscr{R}_h$ are $\big\{I_i\times
J_{i,j}|i\in[l_0],j\in[l_i]\big\}$ and the bitsizes of the
endpoints of $I_i$ and $J_{i,j}$ are bounded by $\OO(d^7\tau)$ and $\OO(d^{11}\tau)$ respectively.

Now it turns to the processes of lifting and computing $s$, both complexity are $\O(d^{19}\tau)$ by Lemma \ref{liftingThm} and Lemma \ref{computingsi}.

Then we choose a rational number (usually integer) $s\in S$ and consider
the sheared space curve $\overline{\C}$, obviously, the bitsize of
$s$ is $\log(d^4\cdot d^{4})=8\log(d)$ according to Lemma 24 of \cite{GPC}. Moreover, the polynomials
$\overline{F},\overline{G}$ have magnitude of $(d,\,d\log d+\tau)$. Then we compute
$\bar{h}=\mbox{sqrfree}(\res_z(\overline{F},\overline{G}))$ which has
magnitude of $(d^2,\OO(d\tau+d^2\log d))$. This resultant computation needs
$\O(d(d+d)^{2+1}d^2(d\log d+\tau))=\O(d^6(d+\tau))$ bit complexity. Then we analyse the topology of the plane curve $\C_{\bar{h}}$ which yields
$$\tilde{\mathcal
{O}}((d^2)^9(d^2+\OO(d^2\log d+d\tau)))=\tilde{\mathcal
{O}}(d^{19}(d+\tau))$$
bit complexity Theorem \ref{plane_top}, this finishes the certifying step.

Due to Lemma \ref{Connectionthm}, the complexity of connection step is also bounded by $\O(d^{19}\tau)$.

In conclusion, the total complexity for all steps are bounded by $\O(N^{20})$, where $N=\max\{d,\tau\}$. Thus, we prove the theorem.
\qed

%

\section*{Acknowledgement} The work is partially supported by NKBRPC
(2011CB302400), NSFC Grants (11001258, 60821002), SRF for ROCS, SEM.

\end{document}